\documentclass[twocolumn,showpacs,pre]{revtex4}
\usepackage{amsmath}
\usepackage{amsfonts}
\usepackage{xspace}
\usepackage{graphicx}

\newcommand{\dd}{\ensuremath{\mathrm{d}}}
\newcommand{\id}{\ensuremath{1 \! \! \mathbb{I}}}
\newcommand{\eg}{e.g.,\xspace}

\newcommand{\ie}{i.e.,\xspace}
\newcommand{\etal}{\emph{et al.}\xspace}
\newcommand{\cf}{cf.\xspace}
\newcommand{\invitro}{\emph{in vitro}\xspace}
\newcommand{\Invitro}{\emph{In vitro}\xspace}
\newcommand{\via}{\emph{via}\xspace}
\newcommand{\beq}{\begin{equation}}
\newcommand{\eeq}{\end{equation}}
\newcommand{\bea}{\begin{eqnarray}}
\newcommand{\eea}{\end{eqnarray}}

\newcommand{\ignore}[1]{}

\begin{document}

\title{Modeling tumor cell migration: From microscopic to macroscopic 
models}

\author{Christophe Deroulers}
\email{deroulers@imnc.in2p3.fr}
\affiliation{IMNC, Universit\'es Paris VII--Paris XI--CNRS, UMR 8165, B\^at. 104, 91406 Orsay Cedex, France}

\author{Marine Aubert}
\email{aubert@imnc.in2p3.fr}
\affiliation{IMNC, Universit\'es Paris VII--Paris XI--CNRS, UMR 8165, B\^at. 104, 91406 Orsay Cedex, France}

\author{Mathilde Badoual}
\email{badoual@imnc.in2p3.fr}
\affiliation{IMNC, Universit\'es Paris VII--Paris XI--CNRS, UMR 8165, B\^at. 104, 91406 Orsay Cedex, France}

\author{Basil Grammaticos}
\email{grammaticos@univ-paris-diderot.fr}
\affiliation{IMNC, Universit\'es Paris VII--Paris XI--CNRS, UMR 8165, B\^at. 104, 91406 Orsay Cedex, France}

\date{\today}

\begin{abstract}
 It has been shown experimentally that contact interactions may 
influence the migration of cancer cells. Previous works have modelized 
this thanks to stochastic, discrete models (cellular automata) at the 
cell level. However, for the study of the growth of real-size tumors 
with several million cells, it is best to use a macroscopic model having 
the form of a partial differential equation (PDE) for the density of 
cells. The difficulty is to predict the effect, at the macroscopic 
scale, of contact interactions that take place at the microscopic scale. 
To address this, we use a multiscale approach: starting from a very 
simple, yet experimentally validated, microscopic model of migration 
with contact interactions, we derive a macroscopic model. We show that a 
diffusion equation arises, as is often postulated in the field of glioma 
modeling, but it is nonlinear because of the interactions. We give the 
explicit dependence of diffusivity on the cell density and on a 
parameter governing cell-cell interactions. We discuss in detail the 
conditions of validity of the approximations used in the derivation, and 
we compare analytic results from our PDE to numerical simulations and to 
some \invitro experiments. We notice that the family of microscopic 
models we started from includes as special cases some kinetically 
constrained models that were introduced for the study of the physics of 
glasses, supercooled liquids, and jamming systems.

\end{abstract}


\maketitle

\section{Introduction}
 The migration of tumor cells plays a principal role in tumor 
malignancy, particularly for brain tumors. The fact that glioma cells 
migrate fast contributes greatly to glioblastoma lethality and thwarts 
therapy strategies based on tumor resection~\cite{giese-invasion}. 
Understanding the mechanisms of migration can be of utmost importance 
when it comes to devising efficient 
treatments~\cite{demuth-review-invasion, nakada-demuth-review-invasion}. 
\Invitro studies combined with mathematical models constitute a 
promising exploration path, a fact supported by several high-quality 
results~\cite{frieboes-model-spheroid-invasion, 
stein-demuth-spheroid-invasion-model}.

 How does one go about modeling cell migration? There are essentially 
two approaches: on one hand, ``macroscopic'' models that take the 
mathematical form of one or several partial differential equations 
(PDE)~\cite{murray-livre-tome-2, 
bellomo-li-maini-revue-modelisation-du-cancer}, and, on the other hand, 
``microscopic'' models where individual cells are represented and evolve 
according to some (stochastic) rules: the so-called individual based 
models, agent based models, cellular automata, and so 
on~\cite{alber-kiskowski-glazier-jiang-revue-automates-cellulaires, 
deutsch-dormann-livre, 
hatzikirou-deutsch-cellular-automaton-migration-review, 
anderson-chaplain-rejniak-livre}.

 In a macroscopic model, one eschews all reference to the elementary 
constituent, the cell, and introduces macroscopic quantities, like the 
average cell density, which evolve according to a PDE. One expects 
stochastic deviations from the average to be negligible at the scale of 
a large population of cells, provided this population is homogeneous 
(quite often, cell populations in cancer are made of cells having 
distinct phenotypes or genotypes, because \eg of genetic 
mutations~\cite{nowell-tumor-progression, 
weinberg-livre-biology-of-cancer, bellomo-delitala-mutations}, and this 
should be taken into account in a more general model of cancer 
development). As a matter of fact, the first brain tumor models (and the 
vast majority of models existing today, which are essentially 
refinements of these first ones) consider the dynamics of a small number 
of macroscopic 
quantities~\cite{tracqui1,tracqui2,woodward-modele-de-base}. The tumor 
cell density is, of course, incontrovertible, the remaining quantities 
being usually concentrations of chemicals, which, one surmises, play a 
role in the migration.

 Using a PDE has many advantages over a ``microscopic'' model. First, 
even with the ultrasimplification of a cellular automaton, one can treat 
a relatively small number of cells, at best a few thousand, a number 
very far from the millions one encounters in a tumor lump. Then, there 
are ready-to-use PDE solvers. Since the PDEs are deterministic, one 
simulation is enough where one should perform many simulations of the 
cellular automaton to average over the stochastic noise. The number of 
discretization steps of the space is fixed only by the mesoscopic 
geometry of the medium in which the cells diffuse; there is no need to 
have as many steps as cells. Finally, it is easier to get analytical 
results from a PDE (even approximate) than from a cellular automaton.

 In this spirit, the simplest approach to cell migration is to 
assimilate the process to linear diffusion. Then a diffusion equation 
may be used to find how the cell density evolves with time, \beq 
\label{eq:edp-diffusion-lineaire} \frac{\partial\rho}{\partial t}= D \, 
\nabla ^2\rho. \eeq This was, for instance, the approach of 
Burgess~\emph{et al.}~\cite{burgess}, who modeled the evolution of 
glioblastomas with the help of a diffusion equation (plus an exponential 
growth representing cell proliferation). Refinements of the diffusion 
model can be, and have been, introduced. One can, for instance, consider 
a space-dependent diffusion coefficient that varies depending on the 
nature of the tissue in which migration takes 
place~\cite{swanson-differential-motility, 
swanson-virtual-and-real-tumors, jbabdi-tensor-diffusion, 
clatz-3D-atlas}. Terms can be added to 
Eq.~(\ref{eq:edp-diffusion-lineaire}) representing various chemotactic 
effects, either as arbitrary external source terms or as quantities that 
couple Eq.~(\ref{eq:edp-diffusion-lineaire}) to other evolution 
equations~\cite{sander-deisboeck-chemotaxis, 
castro-molina-paris-deisboeck-chemotaxis}. When the cells have a 
velocity or a persistence of the direction of motion, one has to study 
(\eg at short time scales of the cell migration) a different type of 
equation, like a hyperbolic PDE or a degenerate parabolic 
PDE~\cite{othmer-hillen-limite-hydrodynamique-diffusive-ii, 
filbet-laurencot-perthame-limite-hydrodynamique-hyperbolique, 
dolak-schmeiser-limite-hydrodynamique-hyperbolique, 
bellomo-et-al-limite-hydodynamique-hyperbolique}.

 However, modeling of cell migration as simple, heatlike diffusion 
(noninteracting random walks~\cite{othmer-et-al-modeles-de-dispersion, 
codling-plank-benhamou-revue-marches-aleatoires-en-biologie}) neglects 
important phenomena, even in the absence of chemotaxis: cells cannot 
penetrate each other, and they often interact \via contacts with other 
cells and/or with the extracellular matrix.

 But what PDE should one use to take contact interactions into account 
in the migration process? It can be easier to devise realistic evolution 
and interaction rules in a cell-based 
model~\cite{schaller-meyer-hermann-comparaison-entre-agent-based-et-pde-pour-spheroides, 
hatzikirou-deutsch-cellular-automaton-migration-review, 
anderson-chaplain-rejniak-livre}. Even though there are situations in 
which it is not possible to discriminate between several macroscopic 
models, especially when only data for the evolution of the cell density 
are at hand~\cite{simpson-et-al-cell-scale-population-scale, 
schaller-meyer-hermann-comparaison-entre-agent-based-et-pde-pour-spheroides, 
cai-et-al-modeles-de-migration}, it is of little interest to use a PDE 
that can later prove to be wrong when more data become available and, in 
particular, when one is able to compare predictions for the trajectories 
of individual cells~\cite{simpson-et-al-cell-scale-population-scale, 
schaller-meyer-hermann-comparaison-entre-agent-based-et-pde-pour-spheroides, 
stein-demuth-spheroid-invasion-model}.

 Sometimes, a microscopic model is derived from a PDE through a 
discretization procedure~\cite{anderson-chaplain-discretisation-edp}. 
But there are several ways to discretize space that lead to very 
different behavior of the microscopic agents, some of which can be 
unrealistic. (For example, consider the discretized population pressure 
model of~\cite{cai-et-al-modeles-de-migration}: one can show that, for 
some discretization procedure, cells move \emph{toward} the high-density 
region they are supposed to flee, but for some other procedure they move 
away from it --- and both correspond to the very same PDE.) At least, 
one needs a biological criterion to choose the ``right'' discretized 
version of the PDE.

 Therefore, to establish a macroscopic model for tumor cell migration 
with contact interactions \via the so-called gap junctions between 
cells, we start from the microscopic model introduced and successfully 
compared to \invitro migration experiments 
in~\cite{aubert-et-al-migration-sans-astrocytes}. In this setting, cell 
proliferation, apoptosis (cell death), and mutations (changes of cell 
behavior) are negligible (hence the population is homogeneous). The 
number of cells is constant and cells simply move on a collagen 
substrate. The model is defined in Sec.~II. (There we also discuss 
special cases in which our model is a kinetically constrained model, 
borrowing some knowledge from the theory of glasses, supercooled 
liquids, and jamming systems.)

 The derivation of the macroscopic model, the so-called hydrodynamic 
limit, is carefully explained in Sec.~III. The derivation of 
hydrodynamic limits is a well-known technique in mathematical 
physics~\cite{lebowitz-presutti-spohn-revue-limite-hydrodynamique, 
spohn-livre-limite-hydrodynamique, 
de-masi-presutti-livre-limite-hydrodynamique, 
kipnis-landim-livre-limite-hydrodynamique, 
deutsch-lawniczak-limite-hydrodynamique}, and it has also been used in 
mathematical 
biology~\cite{bellomo-li-maini-revue-modelisation-du-cancer, 
othmer-hillen-limite-hydrodynamique-diffusive-ii, 
filbet-laurencot-perthame-limite-hydrodynamique-hyperbolique, 
dolak-schmeiser-limite-hydrodynamique-hyperbolique, 
bellomo-et-al-limite-hydodynamique-hyperbolique}. 
See~\cite{painter-discrete-continous-cell-movement, 
drasdo-coarse-graining, alber-et-al-limite-continue-potts-1, 
alber-et-al-limite-continue-potts-2} for examples in the field of cell 
movement and cancer modeling. Since our derivation relies on a 
mean-field like approximation, we review in Sec.~III the circumstances 
under which the approximation fails, and we provide thorough comparisons 
to simulation results.

 The PDE we obtain is a \emph{nonlinear} diffusion equation. Contrary to 
what is customary in the phenomenological models to which we alluded 
above, the terms of the diffusion equation depend crucially on the 
cell-cell interaction present in the microscopic model. As our analysis 
shows, these interactions introduce nonlinearities in the diffusion 
equation. These nonlinearities are essential in order to reproduce the 
density profiles obtained in \invitro migration experiments (Sec.~IV).

\section{The stochastic cellular automaton model}
 In~\cite{aubert-et-al-migration-sans-astrocytes}, a cellular automaton 
(lattice gas) was introduced to mimic the migration of cancer cells with 
contact interactions. The dynamics of migratory motion are described 
through the evolution of points (representing the cells) on a grid (or 
lattice), while the interactions are introduced through evolution rules. 
The impossibility of cells to penetrate each other is trivially taken 
into account (no two cells can share the same position on the grid), and 
the scale length of the lattice step is the typical size of one cell.

 It is easy to extend such a cellular automaton to cells that occupy 
several sites on the lattice, using \eg a cellular Potts model 
~\cite{painter-discrete-continous-cell-movement, 
alber-et-al-limite-continue-potts-1, anderson-chaplain-rejniak-livre, 
alber-et-al-limite-continue-potts-2}. But then we would need to 
introduce many parameters (surface tension, binding energies, ...), 
whereas experimental data at hand were not sufficient to measure so many 
values and avoid dubious fits. Thus the model 
of~\cite{aubert-et-al-migration-sans-astrocytes} is an elementary 
cellular automaton with nonextended cells, involving a single parameter 
that takes contact interactions into account. This approach has made it 
possible to model with success the migration of glioma cells over 
substrates of collagen or of astrocytes and draw conclusions on the 
existence of homotype and heterotype interactions between the 
cells~\cite{aubert-et-al-migration-sans-astrocytes, 
aubert-et-al-migration-avec-astrocytes}.

 In order to define a cellular automaton model, one needs two things. 
First, one must choose the geometry of the lattice. One may work in one, 
two, or three dimensions and consider elementary cells of any shape and 
size. Second, one must specify the update rules of the automaton. With 
the adequate choice of these rules, one can mimic the interactions 
between the elementary entities (glioma cells 
in~\cite{aubert-et-al-migration-sans-astrocytes}) which are modeled by 
the automaton. In this paper, we shall follow the prescriptions we 
introduced in~\cite{aubert-et-al-migration-sans-astrocytes}. We recall 
them briefly below.

\subsection{The geometry}

The cellular automaton introduced 
in~\cite{aubert-et-al-migration-sans-astrocytes} for the description of 
the glioma cell migration is based upon a regular hexagonal tiling (or, 
equivalently, a triangular lattice). This choice is dictated by the fact 
that the hexagonal tiling is the less anisotropic among all regular 
tilings of the plane. Thus one can have a very simple algorithmic 
definition of the geometry while keeping the constraints due to the 
symmetry of the lattice at a minimum. Moreover, in such a lattice each 
cell is surrounded by six sites, a number sufficiently high to allow a 
certain freedom of motion at each update. A generalization to three 
space dimensions is obtained by considering the face-centered-cubic 
(fcc) lattice or the hexagonal compact (hc) lattice, where each site has 
12 nearest)-neighbor sites. It is known that some difficulties arise in 
three-dimensional lattices as compared to two-dimensional lattices when 
one aims at finding a cellular automaton, the continuous limit of which 
would be the Navier-Stokes 
equation~\cite{frisch-hasslacher-pomeau-lattice-gas-navier-stokes, 
dhumieres-lallemand-frisch-lattice-gas-3d}, but here we have no such 
constraint: instead of starting from a macroscopic equation (such as the 
Navier-Stokes equation) and guessing what cellular automaton would 
mimick it, we start from a given cellular automaton and ask for (an 
approximation to) the corresponding macroscopic equation. The only 
constraint is that the macroscopic equation be a reasonable model of the 
behavior of real cells in experiments. Additionally, it turns out that 
both the fcc and hc lattices, although they do not have the same 
symmetry properties~\footnote{The fcc lattice is invariant under the 
symmetry with respect to a site (operation that transforms every vector 
into its opposite) and the hc lattice is not.}, lead to the same 
macroscopic equation. Therefore, we expect that refinements in the 
choice of the three-dimensional lattice would not significantly change 
our results.

 In each case, we denote by $a$ the distance between the centers of two 
nearest-neighbor sites (lattice step).

\subsection{The stochastic evolution rules}

 Each hexagon can be occupied by at most one cell at a time. As a 
consequence, a cell can move only to a free site. The total number of 
cells is locally conserved (but cells may be introduced into or removed 
from the system on the boundaries). Thus our automaton is a kind of 
lattice gas or ``box and ball'' 
system~\cite{takahashi-matsukidaira-box-and-ball}. Notice that our model 
is \emph{not} a so-called lattice-gas cellular automaton (LGCA), where 
individual particles have both a position and a speed: here the cells 
lose the memory of their speed and previous position after each move. In 
the classification of~\cite{othmer-et-al-modeles-de-dispersion}, it is a 
``position jump process,'' not a ``velocity jump process.''

  At each update of the automaton (\emph{discrete time step}), the 
process of picking a lattice site at random and applying the stochastic 
evolution rules to this site is repeated a number of times equal to the 
total number of sites, so that, on average, each site is updated once at 
each time step~\footnote{It would be equivalent to perform the evolution 
of the sites sequentially in an order that is drawn at random before 
each sweep of the lattice, or one could use Gillespie's 
algorithm~\cite{gillespie}.}. If the site is empty, nothing happens. If 
it is occupied by one cell, the following rule is applied. It is 
inspired by what we believe is the biological mechanism of contact 
interaction. The interaction paremeter $p$ is a fixed, constant number 
between 0 and 1. We pick at random one of the six neighboring sites, the 
\emph{target site}. If the target site is occupied, nothing happens. 
Otherwise, the move from the cell to the target site (leaving its 
departure site empty) is done with probability $p$ if at least one of 
the two common neighboring sites of the departure site and of the target 
site is occupied, and with probability $1-p$ if the two common 
neighboring sites are empty. With these assumptions, $p=1$ means that a 
cell can only move if it is in close contact with at least one other 
cell, and it can only move to a position where it will stay in contact 
with a least one of its former neighbors, while, for $p=0$, a cell moves 
only in such a way that all contacts to its former neighbors are broken. 
If $p=\frac{1}{2}$, cells are indifferent to their environment (apart 
from mutual exclusion), and we expect that the cell density obeys a 
diffusion equation. Values of $p>\frac{1}{2}$ model situations in which 
cells are reluctant to break established junctions with their neighbors, 
whereas $p<\frac{1}{2}$ involves a kind of short-range repulsion but 
only between cells that were in close contact, unlike what is usually 
called repulsion between, \eg electric charges. If these rules seem odd 
from a physical point of view (they cannot be related to some energy or 
free energy and convey a strange notion of repulsion, and in particular 
they do not reduce to a special case of the model 
of~\cite{alber-et-al-limite-continue-potts-2}), they incorporate the 
biologically reasonable assumption that a cell makes its decision 
regarding where it goes by considering only information it can get from 
its immediate vicinity, \eg through established communication junctions 
with the cells with which it is in contact.

 The preceding rules can readily be extended to the three-dimensional 
hexagonal compact or fcc lattices; the only change is that a pair of a 
departure and a target site has four neighbors in common instead of two.

 The setting considered in~\cite{aubert-et-al-migration-sans-astrocytes} 
consists in a central part of the lattice occupied by the equivalent of 
the glioma cell spheroid, which, we assumed, acts as a cell source. Once 
a free position in the hexagons surrounding the center appears, it is 
immediately occupied by a cell ``released from the center.'' 
In~\cite{aubert-et-al-migration-sans-astrocytes}, we have explained how 
one can (and must) calibrate the model so as to allow quantitative 
comparisons with experiment.

\subsection{Kinetically constrained models}

 Our cellular automaton contains as special cases (for $p=1$ and 0) two 
kinetically constrained models (KCMs). These models were introduced in 
statistical physics to mimic the dynamics of glasses, supercooled 
liquids, and jamming 
systems~\cite{ritort-sollich-revue-modeles-a-contraintes-cinetiques, 
toninelli-biroli-revue-modeles-a-contraintes-cinetiques}.

 In a KCM, a microscopic object (in our context, cells) can move to a 
free position only if some requirement about its neighborhood is 
fullfilled (\eg only if it has at least another empty position around 
itself). As a consequence, at least for some concentrations of 
surrounding cells, the motion of an individual cell is not a Brownian 
motion, as it is in a diffusive model such as the simple symmetric exclusion 
processes ($p$ close to $\frac{1}{2}$). However, if one does a ``zoom out'' or 
coarse-graining of the trajectory of an individual cell, using 
characteristic length and time scales (which depend on the concentration 
of cells) as new unit length and time, the result looks 
Brownian~\cite{berthier-chandler-garrahan-echelle-pour-loi-de-fick}. 
Above these scales, a diffusion equation should hold.

 The origin of this separation of scales and the existence of the 
characteristic length and time scales is easy to understand for the 
so-called \emph{noncooperative} KCMs (modeling strong glass-former 
materials). There, cells are blocked unless they are close to some 
model-dependent \emph{defect} or \emph{excitation} that has a finite 
size. These excitations travel across the system in a kind of diffusive 
motion, and the configuration of the cells in a region will be changed 
or relaxed only after the region has been traversed by an excitation. 
Even though the excitations, and the individual cells while they are 
close to an excitation, diffuse ``quickly,'' a cell goes ``slowly'' away 
from its initial position because its trajectory is an alternation of 
long nonmoving phases (when the cell is away from any excitation) and 
short diffusive phases (when the cell is close to an 
excitation)~\cite{jung-garrahan-chandler-violation-stokes-einstein, 
berthier-chandler-garrahan-echelle-pour-loi-de-fick}. The characteristic 
time scale is the typical time between two stays of an excitation in the 
neighborhood of a cell, and the characteristic length is the typical 
size of the region explored by a cell during such a stay. Over times 
much greater than the characteristic time, the trajectory of a cell is 
made of many nearly independent non-moving/quickly diffusive patterns 
and, according to the central limit theorem, the cell undergoes a 
Brownian motion with an effective diffusion constant. Fick's law is 
obeyed and the coarse-grained model is like a lattice gas without 
kinetic 
constraints~\cite{toninelli-biroli-revue-modeles-a-contraintes-cinetiques}.

 In the case of the \emph{cooperative} KCMs (modeling fragile 
glass-formers), no traveling finite-size excitation facilitating the 
motion of cells exists. Instead, the moves of the cells are organized 
hierarchically and form structures that can be infinitely 
large~\cite{hedges-garrahan-2tlg}: some changes of the system require 
the participation of an extensive fraction of the cells, hence the term 
``cooperative''. A complete blocking at a finite concentration of cells 
can even be observed for some 
models~\cite{toninelli-biroli-revue-modeles-a-contraintes-cinetiques}. 
Still, in the absence of such blocking, it is possible to define a 
characteristic time and a characteristic scale above which a diffusive 
behavior is recovered and Fick's law is 
obeyed~\cite{pan-garrahan-chandler-2tlg}.

 In our cellular automaton, both the noncooperative and cooperative 
behaviors take place, for $p=1$ and 0, respectively (and the behavior 
for $p$ close to 0, on the one hand, and for $p$ close to 1, on the 
other, is qualitatively the same as in these two limit situations).

 For $p=1$, our model is a noncooperative KCM: according to the rules of 
the automaton, an isolated cell cannot move, but a \emph{cluster} of two 
neighboring cells has a finite probability to go anywhere in the 
lattice and it can be ``used'' to move any isolated cell (it plays the 
role of the aforementioned excitations).

 For $p=0$, our model is a cooperative KCM: this is the very same model 
as the ``2-triangular lattice gas'' or (2)-TLG studied in the context of 
the slow dynamics of glasses~\cite{jaeckle-kroenig-2tlg-autodiffusion, 
kroenig-jaeckle-2tlg-diffusion-collective, pan-garrahan-chandler-2tlg, 
hedges-garrahan-2tlg}.

 To conclude, for general $p$, our model interpolates between a 
cooperative KCM ($p=0$), a simple symmetric exclusion process 
($p=\frac{1}{2}$) and a noncooperative KCM ($p=1$).

\section{The hydrodynamic limit}
 In this section, we explain in detail how we derive an approximate 
macroscopic, deterministic model (which takes the form of a partial 
differential equation) from the microscopic, stochastic cellular 
automaton. This kind of technique has been used in physics since the 
1980s to take the so-called \emph{hydrodynamic limit} of a stochastic 
model. We recall the basics for readers who are not familiar with it. 
Other readers may be interested in our discussion of rigorous results 
and of the quality and limits of our approximation.

 The hydrodynamic limit exists for some model defined on a lattice if, 
as the size of the lattice goes to infinity, one can cut the lattice 
into parts such that (i) the parts are negligibly small with respect to 
the whole lattice, (ii) the values that macroscopic quantities (like the 
cell concentration) take for each part have vanishingly small 
fluctuations, and (iii) the averages of these values obey some partial 
differential equation(s). This process is a kind a coarse-graining. 
Similarly, water, though made of discrete molecules, looks like a fluid 
composed of homogeneous droplets, because each droplet contains a huge 
number of molecules that are typically spread homogeneously in the 
droplet rather than all packed in one-half of the droplet.

 The establishment of a PDE for macroscopic quantities after 
coarse-graining has been extensively studied and can be made 
mathematically rigorous in a number of cases (see, 
\eg~\cite{lebowitz-presutti-spohn-revue-limite-hydrodynamique, 
spohn-livre-limite-hydrodynamique, 
de-masi-presutti-livre-limite-hydrodynamique, 
kipnis-landim-livre-limite-hydrodynamique}). We would like to show that, 
although one might think that this continuous time and continuous space 
approximation is useful only in the case of very large lattices, it 
yields results in excellent agreement with the results from our cellular 
automaton for lattices as small as 16$\times$16. Indeed, even if in one 
simulation (or in one experiment) the microscopic concentration of tumor 
cells is discrete (it can take only two values on each lattice site, 
depending on whether the site is full or empty), the average of this 
concentration over independent simulations of the cellular automaton (or 
over independent experiments) is well predicted by the PDE. Furthermore, 
there are many ways to perform the approximation, some of which being 
rather involved, but we show that, here, an elementary procedure that 
can easily be applied to other cellular automata is sufficient.

\subsection{On the existence and scale of the validity of the 
hydrodynamic limit}

Several authors have proven rigorously that some cellular automata have 
a nonlinear diffusion equation as their macroscopic scaling 
limit~\cite{fritz-preuve-diffusion-non-lineaire-a-partir-de-limite-hydro, 
guo-papanicolaou-varadhan-preuve-diffusion-non-lineaire-a-partir-de-limite-hydro, 
varadhan-preuve-diffusion-non-lineaire-a-partir-de-limite-hydro, 
varadhan-yau-preuve-diffusion-non-lineaire-a-partir-de-limite-hydro}. In 
particular, the theorem 
of~\cite{varadhan-yau-preuve-diffusion-non-lineaire-a-partir-de-limite-hydro} 
proves that our model has a hydrodynamic limit for all values of $p$ 
between 0 and 1 \emph{excluded} (we have to exclude 0 and 1 because, for 
them, some transition rates are 0 and the theorem does not apply). 
However, it seems rather complicated to get an explicit formula for the 
corresponding nonlinear diffusion equation from this theorem. On the 
contrary, our computation provides approximate but explicit formulas 
that are very useful most of the time.

 When $p=1$ or 0, one needs to be more careful: then, we have seen in 
Sec.~II that the cellular automaton is a KCM where Fick's law is obeyed 
only above some characteristic length and time scales.

 Very recently, Gon\c{c}alves~\etal constructed some KCM, the 
hydrodynamic limit of which would be the porous media equations, and 
they were able to prove this 
rigorously~\cite{goncalves-landim-toninelli-porous-medium-equation-comme-limite-hydrodynamique}. 
For $p=1$, our model is only slightly different from theirs and we 
believe that generalizing their proof to ours would be possible. This is 
an indication that (nonlinear) diffusion should hold above \emph{some} 
spatial scale, but in practice it is necessary to know how large this 
scale is. We found that, for $p=1$, and more generally for all values of 
$p$ not close to zero, this scale is of the order of the elementary 
lattice spacing --- the agreement between simulations of the cellular 
automaton and solutions of the PDE is excellent already at this scale 
(see later).

 For $p=0$, we know from the previous section that our model is a 
cooperative KCM (the ``2-triangular lattice 
gas''~\cite{jaeckle-kroenig-2tlg-autodiffusion, 
kroenig-jaeckle-2tlg-diffusion-collective}). To the best of our 
knowledge, there is no rigorous proof of the hydrodynamic limit in this 
case or in a similar case, but it is highly probable that this limit 
exists because the length and time scales we have discussed above exist 
in particular for this model~\cite{pan-garrahan-chandler-2tlg} (they 
diverge as the cell concentration goes to 1, but are otherwise finite). 
We found that our approximation breaks down as $p$ gets close to 0, 
which is a sign that the kinetics cannot be understood only by looking 
at one cell and its nearest neighbors. Still, $p$ needs to be as small 
as 0.05 for one to see a significant discrepancy between our 
approximation and the automaton.

\subsection{Formalism and computation}

 We now proceed with our computation. On the basis of the previous 
considerations, this approximate computation should be valid only if the 
spatial scale above which a macroscopic, coarse-grained diffusive 
behaviour is of the order of the lattice size --- fortunately, this is 
the case for most values of $p$.

 Our procedure involves first taking averages over the stochastic noise, 
neglecting correlations, in order to yield deterministic equations on 
the lattice, then taking the continuous space limit to go from the 
microscopic to the macroscopic scale. Of course, averaging over the 
noise amounts to losing information about the true process. Describing 
the behavior of a few cells on the lattice by the sole average of the 
cell concentration on each lattice site would be a crude approximation, 
of little interest. However, one can hope that this approximation is not 
so bad in the presence of a large number of cells, and that the typical 
fluctuations of interesting macroscopic quantities such as the cell 
concentration become very small as the system grows. Conversely, large 
fluctuations of the macroscopic quantities would become very rare as the 
system grows. As we shall see, this does happen for our model and the 
deterministic, continuous space aproximation is already excellent for 
rather small systems.

\subsubsection{Averaging over the stochastic noise}

 To perform the averaging and neglect correlations, we use a technique 
that is similar to the Hubbard-Stratonovich transform and which was 
already used to study another cellular automaton, the contact 
process~\cite{deroulers-monasson-proccont}. This technique may be used 
to build a perturbative expansion. Here, we will actually stop at the 
first order, since it yields results that are already in excellent 
agreement with our numerical simulations. Therefore, introducing the 
technique is not necessary and one can get the same results using other 
means. But we explain it because its formalism is convenient and 
compact, especially if one wants to use computer algebra to compute the 
macroscopic PDE on complicated or high-dimensional lattices. In 
addition, it yields access to systematic perturbative corrections (we 
can, in principle, get closer and closer to the exact result by keeping 
more and more terms) on one hand, and to large deviations properties, 
on the other hand (see~\cite{deroulers-monasson-proccont} for an example 
where large deviations properties are crucial).

 Using the now standard ``quantumlike'' formalism for master equations, 
introduced a long time ago~\cite{kadanoff-equation-maitresse-quantique, 
felderhof, felderhof-correction, doi}, we start by representing the 
probability distributions of the configurations of the cellular 
automaton as vectors in a simplex. The stochastic rules are encoded as a 
linear operator acting on this space.

 For each site $i$ of the lattice, we introduce two basis vectors 
$|0_i\rangle$ and $|1_i\rangle$ to encode the situations in which the 
site is empty and occupied by one cell, respectively. If the lattice had 
only one site, say $i$, only two configurations would be possible (``the 
site is occupied by a cell'' or ``the site is empty'') and it would be 
enough to specify the occupation probability $\rho$ of the (unique) site 
to characterize entirely the statistical distribution of the 
configurations under some fixed environment. In our formalism, we 
represent this distribution by the vector $v_i := \rho \, |1_i\rangle + 
(1-\rho) \, |0_i\rangle$; the coefficient of each basis vector is the 
probability of observing the corresponding configuration (there is no 
need to take the square modulus as in quantum mechanics). If the lattice 
is composed of two sites, say $i$ and $j$, there are four configurations 
and the most general probability distribution can be represented by the 
vector $\alpha \, |0_i,0_j\rangle + \beta \, |0_i,1_j\rangle + \gamma \,
|1_i,0_j\rangle + \delta \, |1_i,1_j\rangle$ with
$\alpha+\beta+\gamma+\delta=1$ (probabilities must sum up to 1). Here, 
the four basis vectors are the tensor products of the basis vectors for 
the site $i$ and of the site $j$: $|1_i,0_j\rangle = |1_i\rangle \otimes
|0_j\rangle$ and so on. In the special case in which the sites $i$ and $j$
are statistically independent (in which case the probability that $i$ is 
occupied \emph{and} $j$ is occupied is the product of the occupation 
probabilities of $i$ and $j$, say $\rho_i$ and $\rho_j$, respectively), 
this vector factorizes as $[(1-\rho_i) |0_i\rangle + \rho_i |1_i\rangle] 
\otimes [(1-\rho_j)|0_j\rangle + \rho_j |1_j\rangle]$. More generally, 
if there are $N$ sites, there are $2^N$ possible configurations and our 
vector space has dimension $2^N$.

 To explain how we encode the stochastic rules, let us take again the 
simple case in which the lattice has only one site. Let us address first 
the stochastic process of ``radioactive decay'': when the site is full, 
it has a constant probability per unit time $\alpha$ to become an empty 
site, and, when it is empty, it remains so. The occupation probability 
$\rho_i(t)$ of the site obeys the differential equation \beq \dd \rho_i 
/ \dd t = -\alpha \, \rho_i(t) . \eeq This equation is called the 
\emph{master equation} or the Chapman-Kolmogorov equation. The vector 
that encodes the probability distribution, $v_i(t)$, obeys \beq \dd 
v_i(t) / \dd t = \alpha (c_i - c^+_i c_i) v_i(t), \eeq where $c_i$ and 
$c^+_i$ are $2 \times 2$ matrices: in the basis $|0\rangle, |1\rangle$, 
\beq c_i = \left( \begin{array}{cc} 0 & 1 \\ 0 & 0 \end{array} \right) 
\qquad \mathrm{and} \qquad c^+_i = \left( \begin{array}{cc} 0 & 0 \\ 1 & 
0 \end{array} \right) . \eeq The symbols for these matrices are borrowed 
from quantum mechanics, since $c$ and $c^+$ behave like annihilation and 
creation matrices, respectively (for so-called hardcore bosons). The 
evolution operator $c_i - c^+_i c_i$ of vector $v_i$ has two terms: the 
first one, $c_i$, lets the probability of the configuration $|0\rangle$ 
increase proportionnally to the probability of the configuration 
$|1\rangle$, and, without the second one, probability conservation would 
not be ensured because the probability of the configuration $|1\rangle$ 
would be unchanged. Let us give a second example in which two sites $i$ 
and $j$ are coupled: consider the stochastic process in which a particle 
is transferred from the site $i$ (if site $i$ is occupied) to the site 
$j$ (if it is free) at rate $\kappa$, $i$ becoming empty and $j$ full. 
The evolution equation of the vector $v$ encoding the probability 
distribution of the 4 configurations of the couple of sites reads \beq 
\dd v / \dd t = \kappa [ c^+_j c_i - (\id-c^+_j c_j) c^+_i c_i ] v(t) .  
\eeq Here we extend the definition of the matrices. For instance, we 
understand as $c_i$ a $4 \times 4$ matrix, acting on the whole space of 
dimension four where $v$ lives, equal to the tensor product of the 
former matrix $c_i$ (which acts on the two-dimensional space of $v_i$) 
and of the identity matrix of the two-dimensional space of $v_j$. $\id$ 
is the identity matrix acting on the whole space. The first term of the 
evolution operator, $c^+_j c_i$, lets the probability of occupation of 
$j$ increase proportionally to the probability that $i$ was full and 
that $j$ was empty. The second term ensures overall probability 
conservation.

 It is easy to generalize this to any stochastic process. If the process 
has several rules that apply in parallel (\eg ``for any couple of 
neighboring sites $i$ and $j$, a particle can jump from $i$ to $j$ if 
$i$ is full and $j$ empty, making $i$ empty and $j$ full''), the 
evolution matrix is the sum of the evolution matrices encoding each 
individual rule [in the example above, it would be the sum over the 
couples of neighboring sites $i$ and $j$ of the matrices $c^+_j c_i - 
(1-c^+_j c_j) c^+_i c_i$]. It is equivalent to defining a process by its 
stochastic rules and by the expression of the evolution matrix of its 
configuration probability vector $v$.

 Applying this technique, we find that our cell migration process has 
the following evolution matrix: \beq \label{eq:expression-operateur-W} 
\hat{W} = \frac{1}{z} \, \sum_{i,j \ \mathrm{ n.n.}} [c^+_j c_i - 
(1-c^+_j c_j) c^+_i c_i ] \ F \left( \{c^+_k c_k\}_{k \ \mathrm{n.n.\ 
of}\ i \ \mathrm{and}\ j} \right) . \eeq ``n.n.\xspace'' stands for 
\emph{nearest neighbors on the lattice}, and $z$ is their number for a 
given site (coordination number, equal to 6 on the hexagonal tiling and 
to 12 on the three-dimensional fcc or hc lattices). $F$ is a polynom of 
two or several variables; for each $i$ and $j$, it is applied to the 
matrices $c^+_k c_k$, $c^+_l c_l$ and so on where $k$, $l$, ... are the 
common neighbors of $i$ and $j$ on the lattice. The expression of $F$ is 
chosen so that, for each possible configuration of the occupation 
numbers $n_k$, $n_l$, ... of the sites $k$, $l$, ..., $F(n_k, n_l, 
\ldots)$ is equal to the transition rate of one cell from site $i$ to 
site $j$ (provided that site $i$ is full and site $j$ is empty). It 
would be easy to study other rules, but, to conform to the model defined 
in the previous section, we took in two dimensions on the hexagonal 
tiling \beq F(n_k, n_l) = p \, (n_k + n_l) \, (3-n_k-n_l)/2 + (1-p) 
(1-n_k) (1-n_l) \eeq which is equal to \beq 
\label{eq:expression-F-pavage-hexagonal} F(n_k, n_l) = p \, (n_k + n_l - 
n_k n_l) + (1-p) (1-n_k) (1-n_l) \eeq since $n_k^2=n_k$ and $n_l^2=n_l$ 
and, in three dimensions on the fcc or hc lattices, \bea 
\label{eq:expression-F-cfc} F(n_k, n_l, n_m, n_n) = p \, s (5 - s) (s^2 
- 5 s + 10) / 24 + & & \nonumber \\ (1-p) (1-n_k) (1-n_l) (1-n_m) 
(1-n_n), & & \eea where $s$ is short for $n_k + n_l + n_m + n_n$. In 
both dimensions, we want $F$ to be equal to $p$ if at least one of the 
nearest neighbors of $i$ and $j$ is full, and to $1-p$ if all are empty. 
The previous expressions are the simplest ones that achieve this.

 We use the technique of Ref.~\cite{deroulers-monasson-proccont} to 
derive from the expression of the evolution operator $\langle \hat{W} 
\rangle$ the mean-field (correlations are neglected) evolution equations 
for the occupation probability $\rho_i(t)$ of each site $i$. There are 
equivalent techniques~\cite{painter-discrete-continous-cell-movement, 
alber-et-al-limite-continue-potts-1, 
simpson-et-al-cell-scale-population-scale} to get the bare mean-field 
evolution equations of $\rho_i(t)$; this one can additionally yield 
results for the large deviations and systematic perturbative 
corrections. The evolution equation of $\rho_i(t)$ in the typical (\ie 
most probable) configurations reads \beq \partial_t \rho_i(t) = - 
\partial_{\psi_i(t)} \langle \hat{W} \rangle, \eeq where $\partial_t$ 
denotes time derivative, $\partial_{\psi_i(t)}$ is the functional 
derivative with respect to $\psi_i$ at time $t$, taken at $\psi_i=0$, 
and $\langle \hat{W} \rangle$ is the average of the evolution operator. 
This average is obtained by replacing each $\id_i$ with $1$, each $c^+_i 
c_i$ with $\rho_i(t)$, each $c_i(t)$ with $\rho_i(t) \exp[\psi_i(t)]$, 
and consistently each $c^+_i(t)$ with $[1-\rho_i(t)] \exp[-\psi_i(t)]$. 
Using the expression of $\hat{W}$, 
Eq.~(\ref{eq:expression-operateur-W}), one gets the evolution equation 
of the typical site occupation probability on the discrete lattice, \beq 
\partial_t \rho_i(t) = \frac{1}{z} \sum_{j \text{ n.n. of\ } i} 
[\rho_j(t)-\rho_i(t)] F \left[ \{ \rho_k(t) \}_{k\ \mathrm{n.n.\ of}\ i 
\ \mathrm{and}\ j} \right] . \eeq

\subsubsection{Continuous space limit}

 So far, we have obtained $N$ evolution equations for the $N$ average 
occupation probabilities $\rho_i(t)$ of the sites from the initial $2^N$ 
evolution equations for the probabilities of the $2^N$ configurations of 
the lattice. We shall replace the $N$ quantities $\rho_i(t)$ by a single 
field $\rho(\vec{r}, t)$ such that $\rho(\vec{r}_i, t) = \rho_i(t)$ at 
all times $t$, $\vec{r}_i$ being the position of the center of site 
number $i$. If the lattice step $a$ is finite, there are infinitely many 
such fields, but to the $a \to 0$ limit there is only one that is 
regular (twice differentiable). In reality, $a$ is of the order of a 
single cell's diameter and it is not vanishing (this is a natural length 
scale in the problem, and it remains relevant after our procedure): 
taking formally the $a \to 0$ limit means actually that one is 
interested in phenomena that take place on a length scale of many 
individual cells, as is standard in statistical mechanics, by 
restricting to spatial variations of the average cell density $\rho$ 
that are slow on the length scale $a$.

 For a given site $i$, we replace each quantity $\rho_j(t) = 
\rho(\vec{r}_j, t)$ with the Taylor expansion of $\rho(\vec{r}_j, t)$ in 
space around $\vec{r}_i$: introducing the unit vector from $i$ to $j$, 
$\vec{u}_{i \to j} := (\vec{r}_j - \vec{r}_i) / a$, we can write \bea 
\rho(\vec{r}_j, t) & = & \rho(\vec{r}_i, t) + a \, \vec{u}_{i \to j} 
\cdot \vec{\nabla} \rho(\vec{r}_i, t) + \nonumber \\ & & (a^2/2) \, 
\vec{u}_{i \to j} \cdot H(\vec{r}_i, t) \vec{u}_{i \to j} + O(a^3) \eea 
where the center dot denotes the usual scalar product, $\vec{\nabla}$ is 
the gradient operator, and $H$ is the Hessian matrix of $\rho$ (here 
taken at the position $\vec{r}_i$ and at time $t$). After this 
substitution, all terms of order~0 in $a$ vanish because there is 
locally conservation of the number of cells (no proliferation nor 
apoptosis), and all terms of order $a$ vanish because the evolution 
rules are symmetric (induce no bias between left and right) and the 
lattice is reflection-invariant. Also, the lattice is 
translation-invariant and the sites are coupled all in the same way (to 
their nearest neighbors), thus each evolution equation yields the same 
relation between the field $\rho$ and its space and time derivatives. 
Since we are interested in the continuous space ($a \to 0$) limit, we 
disregard terms of order~3 and above in $a$, and after some algebra we 
find the evolution equation of the site occupation probability, \beq 
\label{eq:edp-rho} \partial_t \rho(\vec{r}, t) = a^2 \mathrm{div}[ 
D\left(\rho(\vec{r},t)\right) \vec{\nabla} \rho(\vec{r}, t)], \eeq where 
\beq \label{eq:expression-coeff-diffusion-2d} D(\rho) = (1-p)/4 + (2 
p-1) \rho (1-\rho/2)/2 \eeq on the hexagonal tiling in two dimensions 
(2D) and \beq \label{eq:expression-coeff-diffusion-3d} D(\rho) = (1-p)/6 
+ (2 p -1) \rho (4 - 6 \rho + 4 \rho^2 - \rho^3)/6 \eeq on the fcc 
lattice in 3D. For the hc lattice, we find the very same expression of 
$D(\rho)$ as for the fcc lattice. The expression of $D(\rho)$ is the 
value of the function~$F$ [Eqs.~(\ref{eq:expression-F-pavage-hexagonal}) 
and (\ref{eq:expression-F-cfc})] when all its arguments are equal to 
$\rho$, up to a geometrical factor. In the $a \to 0$ limit, the 
right-hand side of Eq.~(\ref{eq:edp-rho}) vanishes, leaving the trivial 
equation~$\partial_t \rho(\vec{r}, t) = 0$: this is a sign that nothing 
interesting happens, in the limit of continuous space, at the time scale 
of the updates of individual sites, $t$. The only time scale leading to 
a nontrivial equation has a unit time of the order of 
$1/a^2$~\footnote{In particular, our cellular automaton has no 
hyperbolic macroscopic 
limit~\cite{filbet-laurencot-perthame-limite-hydrodynamique-hyperbolique, 
bellomo-et-al-limite-hydodynamique-hyperbolique}, because it is a 
``position jump,'' not a ``velocity jump,'' 
process~\cite{othmer-et-al-modeles-de-dispersion}. The possibility of 
the persistence of the direction of motion of the cells was briefly 
discussed and negatively compared to experiments 
in~\cite{aubert-et-al-migration-sans-astrocytes}.}. This scale is 
characteristic of diffusion (in the case of an advective phenomenon, we 
would have chosen the time scale 
$1/a$~\cite{lebowitz-presutti-spohn-revue-limite-hydrodynamique}). 
Equation~(\ref{eq:edp-rho}) is a diffusion equation with a 
concentration-dependent diffusion coefficient $D(\rho)$. This equation 
belongs to the family of the porous media 
equations~\cite{livre-porous-media-equation}. A striking difference with 
the heat equation (or with Fick diffusion) comes from the possibility 
that $D$ vanishes at some site occupation probability $\rho$, which may 
create abrupt fronts in the diffusion profile instead of large, Gaussian 
tails.

 To completely free ourselves from the lattice, which in most practical 
applications does not exist, but was convenient to define the model, we 
write down the equation for the spatial concentration of cells $c$ 
instead of the site occupation probability $\rho$. This is done be 
dividing $\rho$ by the volume of a single site, $\sqrt{3} a^2/2$ for the 
hexagonal tiling, and $a^3/(3 \sqrt{6})$ for the fcc lattice. Because 
$\rho$ and $c$ are proportional, we can discuss either one in the rest 
of this Section; when we discuss experimental results, we shall use only 
$c$.

 Before proceeding to a comparison of this result to numerical 
simulations, let us discuss the vector ``density of cell current,'' 
$\vec{\jmath}$. We expect it to satisfy Fick's law, with a nonlinear 
diffusivity. To compute $\vec{\jmath}$, we first compute the net cell 
current along the lattice link $i \to j$, \ie the current from site $i$ 
to site $j$, \beq \langle c^+_j c_i - c^+_i c_j \rangle = \frac{a}{z} 
F\left(\left\{\rho(\vec{r}, t), \ldots \right\}\right) \vec{u}_{i \to j} 
\cdot \vec{\nabla} \rho (\vec{r}, t) + O(a^3), \eeq where $\vec{r} := 
(\vec{r}_i + \vec{r}_j)/2$ and all arguments of $F$ are equal to 
$\rho(\vec{r}, t)$. Then we add the contributions of all links (on the 
hexagonal tiling, there are three types of contributions since the links 
can have three directions), and we go to the \emph{density} of current 
by multiplying each contribution by the density of links with that 
direction on a hyperplane orthogonal to that direction. On the hexagonal 
tiling, if some links are, say, vertical, then we count the density of 
intersections of a horizontal line with links. These intersections have 
a periodic pattern and can be grouped four by four; a group has the 
length $\sqrt{3} a$ and contains two vertical links (each of these 
contributes one), and two links making an angle $\pi/3$ with the 
vertical direction [each of these contributes $\cos(\pi/3)$], hence the 
factor $\sqrt{3}/a$. We find on the hexagonal tiling \beq 
\label{eq:loi-de-fick-pavage-hexagonal} \vec{\jmath}(\vec{r}, t) = 
-\frac{2}{\sqrt{3}} D[\rho(\vec{r}, t)] \vec{\nabla} \rho(\vec{r}, t) . 
\eeq Putting this expression into the conservation equation for the 
number of cells (there is neither apoptosis nor proliferation in our 
model), \beq \label{eq:conservation-nombre-de-cellules} \partial_t 
c(\vec{r}, t) = - \mathrm{div} \, \vec{\jmath}(\vec{r}, t), \eeq and 
using that $c = 2 \rho / (\sqrt{3} a^2)$ (on the hexagonal tiling), we 
find again Eq.~(\ref{eq:edp-rho}).

\subsection{Comparison to steady-state simulations of the automaton}

\begin{figure} \begin{center}
\includegraphics[width=\linewidth]{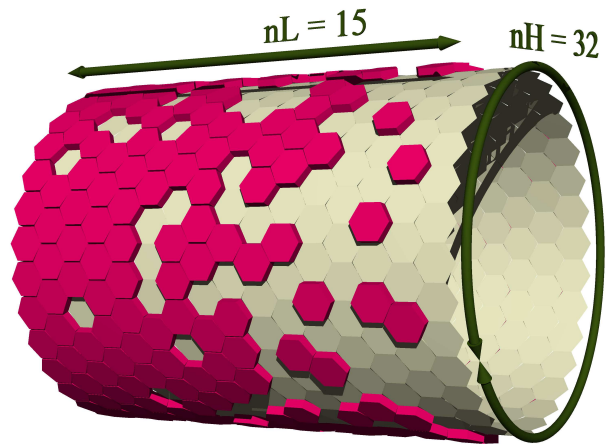}
\caption{\label{fig:cylindre-configuration-typique} \emph{(Color 
online)} 3D view of the first lattice geometry we used to test our 
results: a cylinder (lattice with periodic boundary conditions in one 
direction) connected to a full reservoir of cells (a source) on its left 
end and to an empty reservoir of cells (a sink) on its right end. Here, 
the system is made of $n_L+1=16$ rings of sites (including the two rings 
in the reservoirs), and each ring is made of $n_H=16$ sites. After some 
transient regime, a steady current of cells establishes along the 
cylinder. A typical configuration of the cells in the steady state for 
interaction parameter $p=0.8$ is shown.}
\end{center}
\end{figure}

 In order to test the analytical results~(\ref{eq:edp-rho}) and 
(\ref{eq:expression-coeff-diffusion-2d}), we did some simulation of the 
cellular automaton in a simple geometry. More realistic simulations are 
the topic of the next Section. Since, as will be discussed later on, the 
equilibrium state of the cellular automaton is trivial (all 
configurations of the tumor cells are equally probable), we chose a 
setting where a nonequilibrium steady state of the tumor cells can 
exist: a cylinder made of $n_L+1$ rings of cells (its total length is 
$(n_L + 1) a \sqrt{3}/2$) and base circumference $H = n_H a$ connected 
to a reservoir full of cells ($\rho=1$) at the left end and to an empty 
reservoir ($\rho=0$) at the right end, as shown in 
Fig.~\ref{fig:cylindre-configuration-typique} ($n_L$ and $n_H$ are two 
integers). The first and the last ring of cells of the cylinder belong 
to the full and to the empty reservoirs respectively; hence the 
effective length over which diffusion takes place is $L := n_L a 
\sqrt{3}/2$. We chose a cylinder rather than a simple rectangle in order 
to make the boundary effects in the direction perpendicular to the flow 
small, if not vanishing. We introduce two coordinates on the cylinder: 
$x$ is measured in the direction parallel to the axis and $y$ along the 
shortest circles. To simulate the reservoirs, on the boundaries of the 
system, cells behave under special evolution rules: each time one of the 
sites on the left border of the lattice is left by a cell, it becomes 
occupied again by a new cell with no delay, and each time a cell arrives 
on a site of the right border of the lattice, it is removed at once. In 
this geometry, if we let the system evolve starting from a random 
configuration, it relaxes to a steady state with a permanent current 
where gains and losses of cells from and to the two reservoirs 
compensate in average (but the instantaneous total number of cells still 
fluctuates as time advances). On an infinite lattice, the steady state 
might be reached only after an infinite duration. In practice, we 
simulate in parallel two families of independent systems, one where the 
site occupation probability is initially close to one and one where it 
is initially close to zero, and we stop our simulation when the values 
of the macroscopic observables (averaged over the systems of each 
family) are undistinguishable up to the error bars. The number of 
systems in each family is chosen so that the error bars are shorter than 
the precision that we request in advance.

 In the steady state, the concentration profile can be computed in the 
macroscopic limit thanks to Fick's 
law~(\ref{eq:loi-de-fick-pavage-hexagonal}) or to the 
PDE~(\ref{eq:edp-rho}). In the stationary regime (where no macroscopic 
quantity depends on time), any initial asymmetry has been ``forgotten'' 
by the system and $\rho(\vec{r})$ does not depend on $y$ because the 
system is translation-invariant in the $y$ direction. Furthermore, 
$\rho(\vec{r})$ is such that the current of cells is uniform: 
$\vec{\jmath}$, the density of cell current, is independent of $x$ and 
$y$. Then Fick's law~(\ref{eq:loi-de-fick-pavage-hexagonal}) reads \beq 
(2/\sqrt{3}) D[\rho(x)] \, \partial_x \rho(x) = - j. \eeq With the 
boundary conditions $\rho(0)=1$ and $\rho(L)=0$ and the expression for 
$D(\rho)$ above, we find in 2D \beq 
\label{eq:densite_de_courant_j_stationnaire} j = (1+p)/(6 \sqrt{3} L) 
\eeq and \beq \label{eq:densite_stationnaire_2d} x(\rho) = \{1 + \rho 
[(2 p - 1) \rho (\rho - 3) - 3 (1-p)]/(p+1) \} L. \eeq Then one can get 
the curve for $\rho(x)$ by plotting the parametric curve $(x(\rho), 
\rho)$.

\begin{figure} \begin{center}
\includegraphics[width=0.8\linewidth]{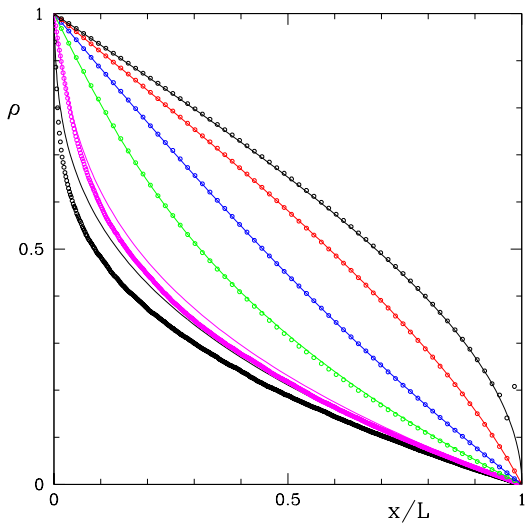} 
\caption{\emph{(Color online)} Dependence on the interaction parameter 
$p$ of the stationary profiles $\rho(x/L)$ of the site occupation 
probability along the cylinder between the full reservoir (left, $x=0$, 
site occupation probability $\rho=1$) and the empty reservoir (right, 
$x=L$, $\rho=0$) --- see 
Fig.~(\ref{fig:cylindre-configuration-typique}). Since the cell 
concentration $c$ is proportional to the site occupation probability 
$\rho$, the profiles of $c$ would be the same. For each value of~$p$, we 
plot the simulation results for the cellular automaton (circles) and our 
prediction from the analytic approximation 
Eq.~(\ref{eq:densite_stationnaire_2d}) (solid lines). From bottom to 
top, $p=0.01$, 0.05, 0.2, 0.4, 0.7 and 1. The error bars are smaller 
than the circles' size. The cylinder is made of $64 \times 64$ sites 
except for $p=0.01$ and 0.05 where there are $512 \times 512$ sites.}
\label{fig:comparaison-densite-stationnaire-2d}
\end{center}
\end{figure}

\begin{figure} \begin{center}
\includegraphics[width=0.8\linewidth]{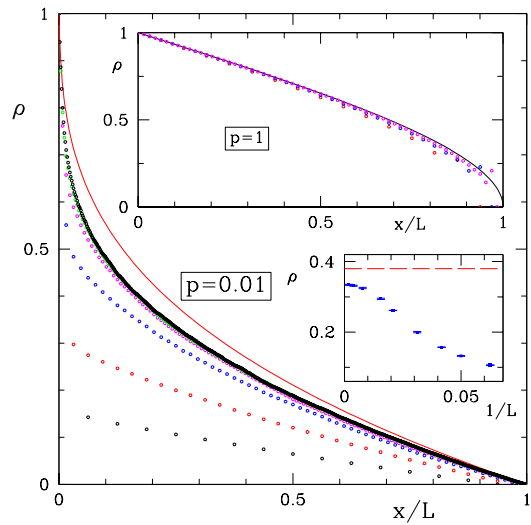}
\caption{\emph{(Color online)} Finite-size effects for the stationary 
profile of the cell concentration between the two reservoirs. 
\textbf{Main curve:} dots: profile $\rho(x/L)$ of the site occupation 
probability from simulation results for $p=0.01$ and system sizes 
$n_L=n_H-1=15$, 31, 63, 127, 255, and 511 (from bottom to top). It shows 
a strong finite-size dependence (for $p=0.05$, not shown, a weaker 
finite-size dependence is found). The solid line is the analytic 
approximation, clearly in disagreement with the simulation results. 
\textbf{Bottom inset:} for p=0.01, the extrapolation of the occupation 
probability $\rho$ at position $x=L/4$ to an infinite lattice shows that 
the discrepancy between the simulation data (dots) and the analytical 
approximation (dashed line) is not due to finite-size effects. Indeed, 
there is convergence toward a value of the site occupation probability 
that is clearly below the analytic approximation. \textbf{Top inset:} 
profile $\rho(x/L)$ for $p=1$, with numerical data for the sizes 
$n_L=15$, 31, and 63 only (dots) and analytic prediction (solid line). 
The finite-size effects are very weak and there is a good agreement 
between simulation data for large systems and the analytic prediction. 
This is true more generally for all values of $p$ away from 0 (say, 
larger than 0.1).}
\label{fig:comparaison-densite-stationnaire-2d-etf} 
\end{center}
\end{figure}

\begin{figure} \begin{center}
\includegraphics[width=0.8\linewidth]{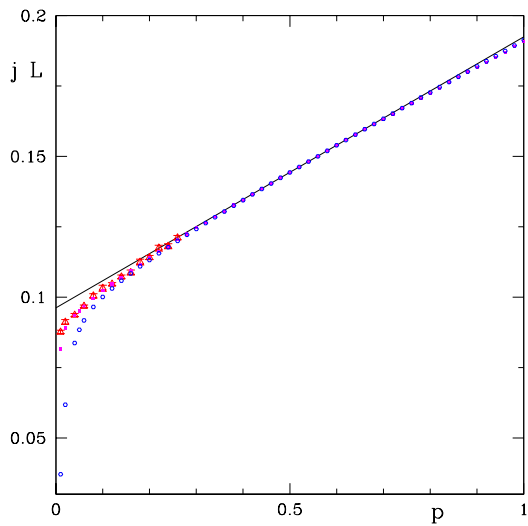} 
\caption{\emph{(Color online)} Stationary current of cell density 
between the full and the empty reservoirs, across the cylinder (geometry 
defined in Fig.~\ref{fig:cylindre-configuration-typique}), as a function 
of the interaction parameter $p$. To be able to compare different system 
sizes (bottom to top~: $nH-1=n_L=15$, 63 and 255), we actually plot the 
current density $j$ multiplied by the system length $L$ (dots). For 
$n_L=255$, only the points that differ significantly from the other 
sizes are represented. The error bars are smaller than the symbol's 
sizes, except for $n_L=255$, where they are drawn. Solid line: analytic 
approximation, Eq.~(\ref{eq:densite_de_courant_j_stationnaire}) times 
$L$. Like for the profile of the cell concentration, there is a large 
disagreement between the analytic approximation and the simulation 
results when $p$ is close to 0 and a small disagreement when $p$ is 
close to 1.}
\label{fig:comparaison-courant-stationnaire-2d}
\end{center}
\end{figure}

 In Fig.~\ref{fig:comparaison-densite-stationnaire-2d}, we plotted the 
stationary profile of the site occupation probability cell $\rho$ on the 
cylinder, between the two reservoirs, as obtained from simulations of 
the cellular automaton and from the macroscopic PDE. Notice the effect 
of nonlinear diffusion: for $p=1/2$ (not shown in 
Fig.~\ref{fig:comparaison-densite-stationnaire-2d}), the profile is 
linear as predicted by the usual Fourier or Fick law, but, for other 
values of $p$, interactions between cells make it more complicated. For 
most values of $p$, there is an excellent agreement between the 
simulation results and the analytical approximation for all lattice 
sizes we tried (starting from $L=16$). However, we can distinguish two 
situations in which the microscopic and the macroscopic models disagree. 
The first one is related to a boundary effect: at the vicinity of the 
empty reservoir when $p=1$ (respectively at the vicinity of the full 
reservoir when $p=0$), a clear discrepancy of the average occupation 
probability $\rho$ of the last-but-one lattice site (respectively of the 
first few lattice sites) is seen between the cellular automaton results 
and the corresponding value from our analytic formula. The second 
situation is what happens when $p$ gets close to 0; see, \eg $p=0.01$ 
and $0.05$ in Fig.~\ref{fig:comparaison-densite-stationnaire-2d}. Here, 
the concentration profiles of the finite cellular automata show a strong 
finite-size dependence (main plot of 
Fig.~\ref{fig:comparaison-densite-stationnaire-2d-etf}, compared to the 
top inset of Fig.~\ref{fig:comparaison-densite-stationnaire-2d-etf}), 
and they seem to converge to some shape that is clearly different from 
the profile predicted from the PDE. To exclude the possibility that the 
simulation results agree with the analytic approximation at very large 
sizes, but not at the sizes we have simulated, we plot in the bottom 
inset of Fig.~\ref{fig:comparaison-densite-stationnaire-2d-etf} an 
extrapolation to infinite systems: it shows that the discrepancy should 
persist even for huge systems. These phenomena are confirmed in 
Fig.~\ref{fig:comparaison-courant-stationnaire-2d}, where the permanent 
current of cells across the cylinder is plotted as a function of $p$: 
one sees a good agreement for most values of $p$, with a slight 
deviation and practically no finite-size effects around $p=1$, and some 
more important deviation, together with strong finite-size effects, 
around $p=0$. This will be understood in the next subsection.

\subsection{Discussion of our approximations: The occurrence of 
correlations}

 Our approximate computation of the hydrodynamic limit fails if the 
correlations between the occupation of neighboring sites are not 
negligible or if the continuous space limit has singularities (\eg an 
infinite gradient) that cannot exist on the lattice.

\subsubsection{Singularities in continuous space}

 Singularities are easy to predict and to address; they could happen for 
our model since the diffusion coefficient vanishes (for $p=1$ and 
$\rho=0$ on one hand and for $p=0$ and $\rho=1$ on the other). Indeed, 
Fick's law reads \beq \vec{\jmath}(\vec{r}, t) = -D[\rho(\vec{r},t)] 
\vec{\nabla} \rho(\vec{r},t), \eeq and at a point where $D$ vanishes 
while $\vec{\jmath}$ is finite (which happens for $p=1$ or 0 between 
the two reservoirs of the previous subsection), the concentration gradient 
is infinite. To get the correct, physical behavior if necessary, it may 
suffice to reintroduce two or a few lattice sites around the point where 
the singularity appears and to solve separately the discrete equations 
(obtained after averaging but before the Taylor expansion) on these 
sites, and the PDE on the rest of the lattice.

 In our case, this simple procedure does \emph{not} enable one to 
quantitatively (or even qualitatively) understand the discrepancies 
between the automaton and the approximate results close to the 
reservoirs. This is because these discrepancies have to do with 
correlations.

\subsubsection{Correlations}

 If $i$ and $j$ are two sites, let us consider the so-called connected 
correlation function~\footnote{This name comes from the fact that these 
correlation functions have connected diagrams in a diagrammatic 
expansion of all correlation functions --- see, 
\eg~\cite{hansen-macdonald-livre}.} of $i$ and $j$, \bea \langle 
\rho_i(t) \rho_j(t) \rangle_c & := & \langle [\rho_i(t) - \langle 
\rho_i(t) \rangle ] [ \rho_j(t) - \langle \rho_j(t) \rangle ] \rangle 
\nonumber \\ & = & \langle \rho_i(t) \rho_j(t) \rangle - \langle 
\rho_i(t) \rangle \langle \rho_j(t) \rangle . \eea This quantity is zero 
if the statistical distribution of the occupation numbers of the sites 
$i$ and $j$ is the same as if the sites $i$ and $j$ would be filled 
independently at random with given mean occupation probabilities. More 
precisely, in the case of a finite lattice of $N$ sites with a fixed 
total number of cells, this definition must be replaced with \beq 
\langle \rho_i(t) \rho_j(t) \rangle_c := \langle [\rho_i(t) - \langle 
\rho_i(t) \rangle ] [ \rho_j(t) - \langle N \rho_j(t) - 1 \rangle/(N-1)] 
\rangle \eeq if one wants to detect correlations due to the interactions 
between cells that could be ``hidden'' behind the contribution 
$-\rho(1-\rho)/(N-1)$ due to the constraint that the total number of 
cells is fixed.

 Our approximate computation of the hydrodynamic limit makes use of an 
assumption of statistical independence (``all connected correlations 
functions are zero''), and we found in our simulations that the 
correlations are indeed small (they vanish as $L \to +\infty$) in almost 
all cases.

 A situation in which we can compute exactly the correlations is the
equilibrium state of a translation invariant lattice. Since, in our
stochastic automaton, any movement of a cell may be reversed and since,
if the lattice is translation invariant, any transition from one
configuration of the cells on the lattice to another has the same rate,
$1/(6N)$, the detailed balance
condition~\cite{ritort-sollich-revue-modeles-a-contraintes-cinetiques}
is fulfilled, the equilibrium state exists, and the probability
distribution of the configurations is uniform. Then the connected
correlation functions (with the definition above, relevant to the case
of conserved total cell number) between any two sites are zero.  We
simulated the cellular automaton in this situation and we checked that
our numerical data are compatible with vanishing correlations (up to the
error bars) for several values of $p$ and of the cell concentration;
this is a check of correctness of the simulation program.  In this case,
our analytical approximation is exact, but the result is of course
trivial ($\rho$ is uniform).

\begin{figure} \begin{center}
\includegraphics[width=0.623\linewidth]{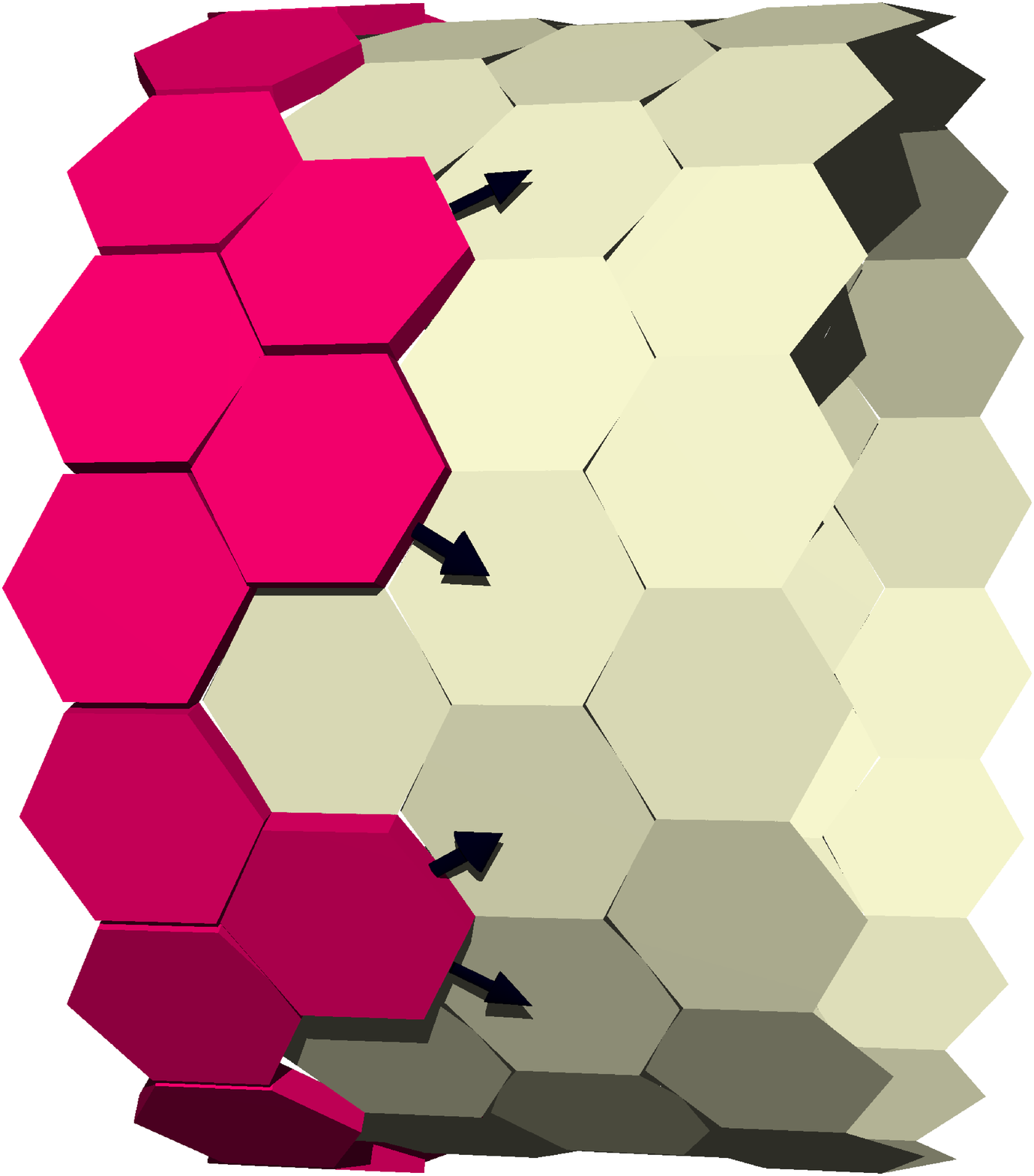}%
\includegraphics[width=0.367\linewidth]{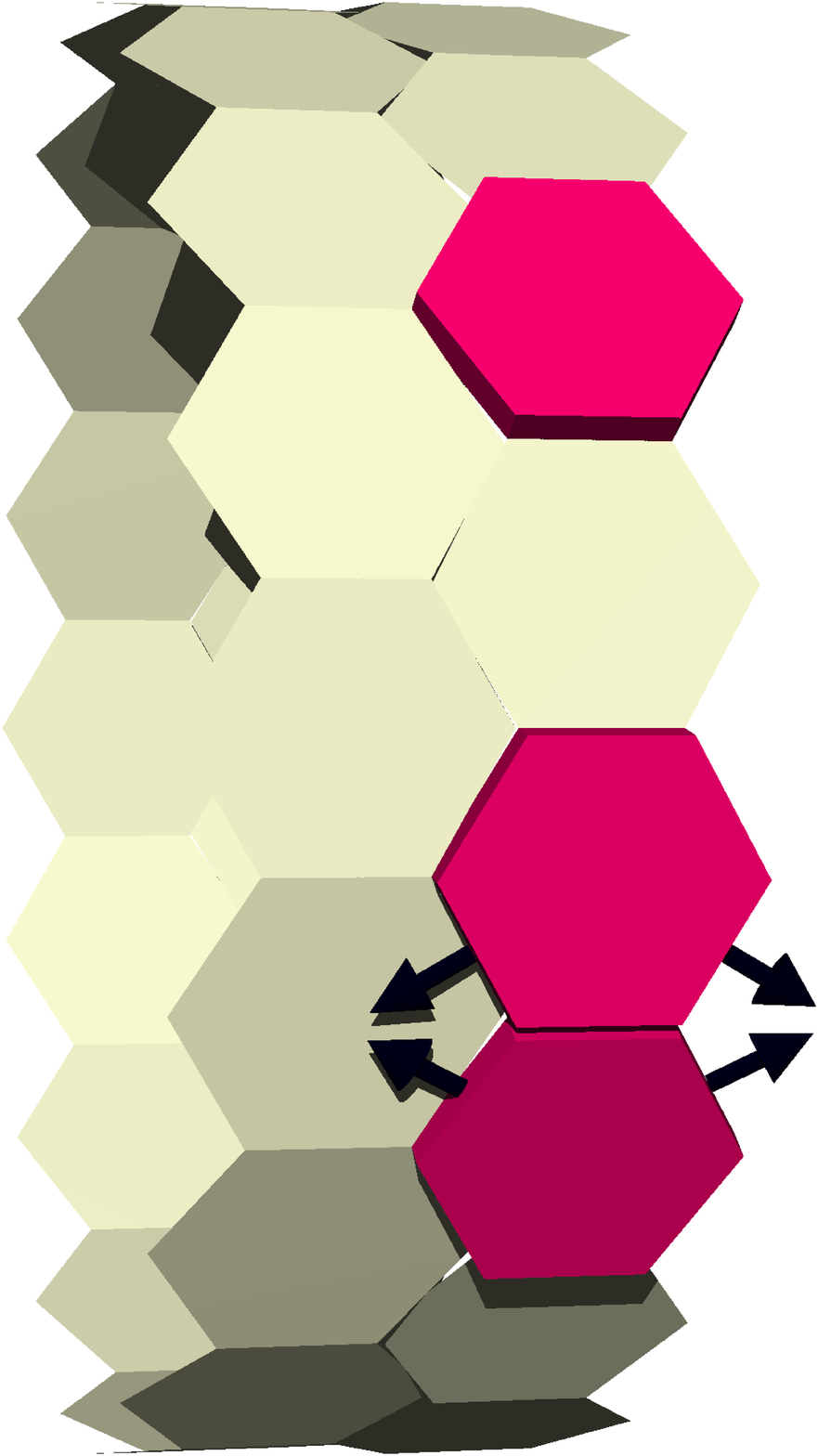}
\caption{\emph{(Color online)} Schematic explanation of the correlations 
that appear for $p$ close to 0 near the full reservoir (left) and for 
$p$ close to 1 near the empty reservoir (right). Left: when two cells 
sit next to one another close to the full reservoir and if $p$ is close 
to 0, they prevent one another from moving in one of the two directions 
that would be allowed if they were alone (black arrows represent 
permitted moves). Thus accompanied cells stay longer close to the 
reservoir than single ones, and positive correlations occur (see text). 
Right: when one cell sits close to the empty reservoir and if $p$ is 
close to 1, it cannot move unless it has a neighboring cell. Thus single 
cells stay longer close to the empty reservoir than accompanied ones, 
and negative correlations occur (see text).}
\label{fig:explication-correlations-pres-reservoirs}
\end{center}
\end{figure}

 Let us come back to the nonequilibrium steady state between the two
reservoirs. In such a case, it is well known that correlations do exist,
can be quantitatively important, and, above all, are generically
long-range~\cite{spohn-correlations-hors-equilibre,
garrido-lebowitz-maes-spohn-correlations-hors-equilibre}. For our
automaton, we found numerically that, for $p$ close to 1 (0) there
exists a significant negative (positive) correlation of the
occupation probability of adjacent sites in the last (first) row of
sites, parallel to the reservoir. This is why the macroscopic limit
disagrees, in these situations, from the simulation. In other
situations, they decrease at least as fast as $1/L$ as the system length
$L$ goes to $+\infty$. Since studying these correlations in more detail
would go beyond the scope of the present article, we limit ourselves here
to explaining qualitatively what happens close to the boundaries. The
correlation there can be interpreted using the microscopic rules (see
Fig.~\ref{fig:explication-correlations-pres-reservoirs}). If $p=1$,
when two cells are close to the empty, absorbing reservoir
(Fig.~\ref{fig:explication-correlations-pres-reservoirs} right), sooner
or later one of the two cells will fall into the reservoir, leaving the
other one alone. This other cell will not move until another cell comes on
a neighboring site. Therefore, isolated cells stay longer close to the
reservoir than accompanied cells, and there is more chance to find an
empty site close to a cell than one would expect if cells were
distributed at random, hence the negative correlations. Far away from
the reservoir, the cell concentration is higher and there is always some
cell to ``unblock'' an isolated cell, so the same phenomenon cannot be
observed. Conversely, if $p=0$, when two cells are close to the full
reservoir (Fig.~\ref{fig:explication-correlations-pres-reservoirs},
left), they are in some sense in each other's way and they may progress
away from the reservoir only in one of the two directions they could use
if they were alone, the other direction being forbidden by the kinetic
rule of the automaton since they would stay in contact with a cell they
were already in contact with. If one could consider only these cells on
the nearest line of sites from the reservoir and forget about the cells
that lie on the next-to-nearest line of sites, one could conclude that
accompanied cells stay longer close to the reservoir than single cells,
that there is more chance to find another cell close to a cell than one
would expect if cells were distributed at random, and that the
correlation is positive. This is not so simple because the
next-to-nearest line of sites has many cells that can compensate this
effect, and we do observe that correlations extend over a few lines of
sites away from the full reservoir, but the qualitative idea is correct. 
Far away from the full reservoir, the site occupation probability is not
close to 1 and there is always some hole to ``unblock'' a cluster of
cells, hence no such correlation is observed. We expect that taking into
account these correlations into the analytic computation would improve
the agremeent between simulations and analytic formulas.

 Now let us come to the disagreement of the simulated and predicted 
concentrations of cells in the whole system when $p$ comes close to 
zero. Our interpretation is the following. We know from a preceding 
discussion that, at $p=0$, the cellular automaton is a cooperative KCM 
for which Fick's law is obeyed only above some spatial scale that 
diverges with the cell concentration. Since, between the two reservoirs, 
all concentrations can be observed, Fick's law might never hold even in 
arbitrarily large systems, but this is beyond the scope of the present 
paper. If $p$ is close to zero but strictly positive, we expect that the 
threshold spatial scale for effectively Brownian diffusion should be 
large (diverging with $p$) but finite at all values of the cell 
concentration. Therefore, and contrary to what happens for larger values 
of $p$ where the threshold scale must be comparable to the lattice 
spacing, there is a whole range of system sizes where the Brownian 
regime is not yet reached and the concentration profile depends strongly 
on the system size. Above this scale, diffusion sets in but, as a 
consequence of the cooperative nature of the kinetics, correlations 
always play a significant role and our evaluation of the diffusion 
constant, which neglects them, is irrelevant, hence the disagreement 
observed even at large system sizes between the predicted profile and 
the actual one.

\section{Results for cells diffusing out of a spheroid}
\newcommand{\Dud}{\ensuremath{D_*}}

In this section, we come back to the radial geometry, where cells 
diffuse away from a center (\cf Sec. II~B), in order to compare the 
results from the macroscopic model to experimental results. We use the 
expression of the diffusion coefficient obtained in Sec. III~B~2, \cf 
Eq.~(\ref{eq:expression-coeff-diffusion-2d}), for a 2D system on a 
hexagonal lattice. We calculate numerically the solution of the 
nonlinear diffusion equation in a radial geometry by using a fully 
implicit method where the spatial derivatives in the diffusion equation 
are evaluated at time step $t+1$ (for the sake of 
stability)~\cite{numerical-recipes}. The site occupation probability 
$\rho$ is constrained to be equal to $1$ in the center at all times.

\begin{figure}[http]
	\begin{center}
		\includegraphics[width=\linewidth]{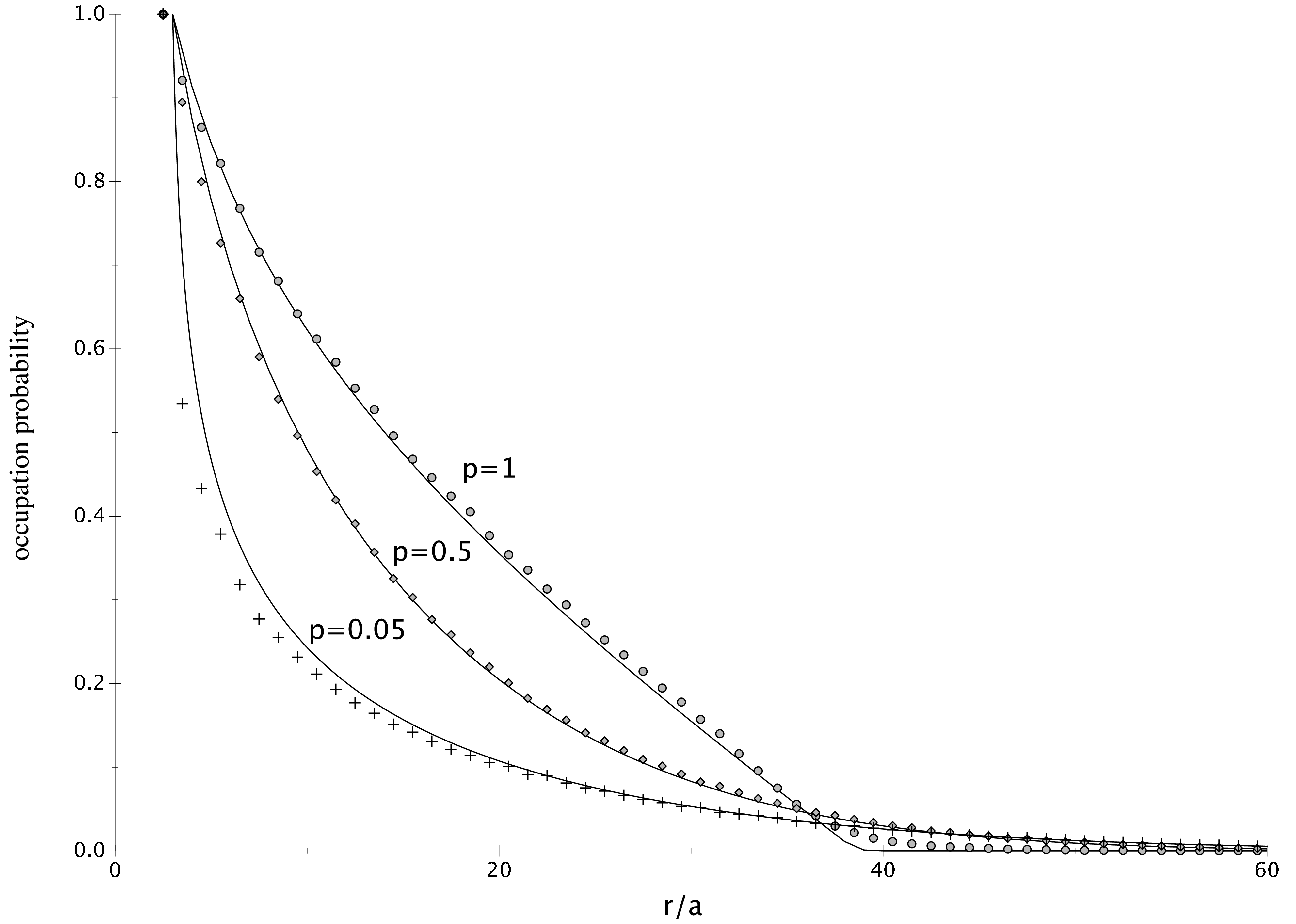}
	\end{center}
	\caption{Density profiles from the automaton (dotted line) and 
from the numerical solution of the diffusion equation (full line) for 
$p=1$, $0.5$, and $0.05$. Error bars for the automaton density profile 
are smaller than the markers and thus are not represented. }
	\label{fig:comp}
\end{figure}

A comparison between the density profiles from the automaton and from 
the numerical solution of the diffusion equation at the same time is 
presented in Fig.~\ref{fig:comp}. As in the case of diffusion between a 
full and an empty reservoir, the agreement is very good except for 
$p=0.05$, close to the full reservoir, where the numerical solution does 
not agree with the density profile of the automaton as nicely as in 
the cases of larger $p$. This effect is due to correlations between 
close neighbors which are not negligible when $p$ is close to 0 (\cf 
Sec. III~D~2). In general, the agreement between both profiles is 
nice. This is a supplementary check of our derivation of a nonlinear 
equation from the automaton.

\indent

\begin{figure}[http]
	\begin{center}
		\includegraphics[width=\linewidth]{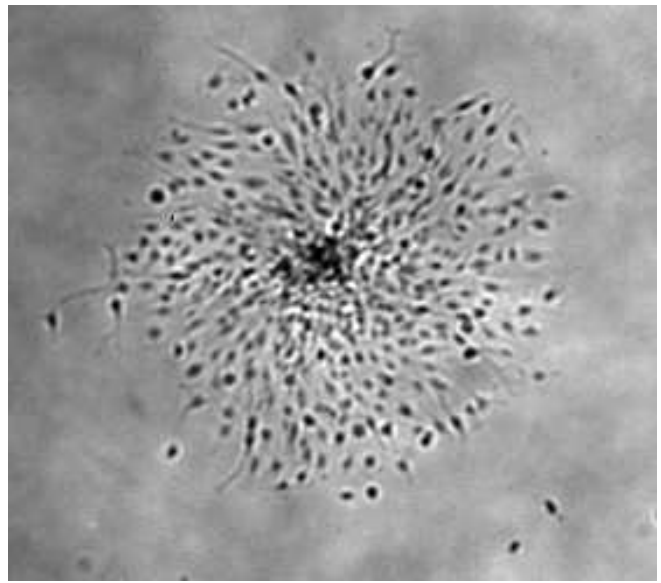}
	\end{center}
	\caption{Experimental pattern of migration after 36~h. The real 
size of the image is $1.90 \times 1.77$~mm.}
	\label{fig:spheroid}
\end{figure}

\begin{figure}[http]
	\begin{center}
		\includegraphics[width=\linewidth]{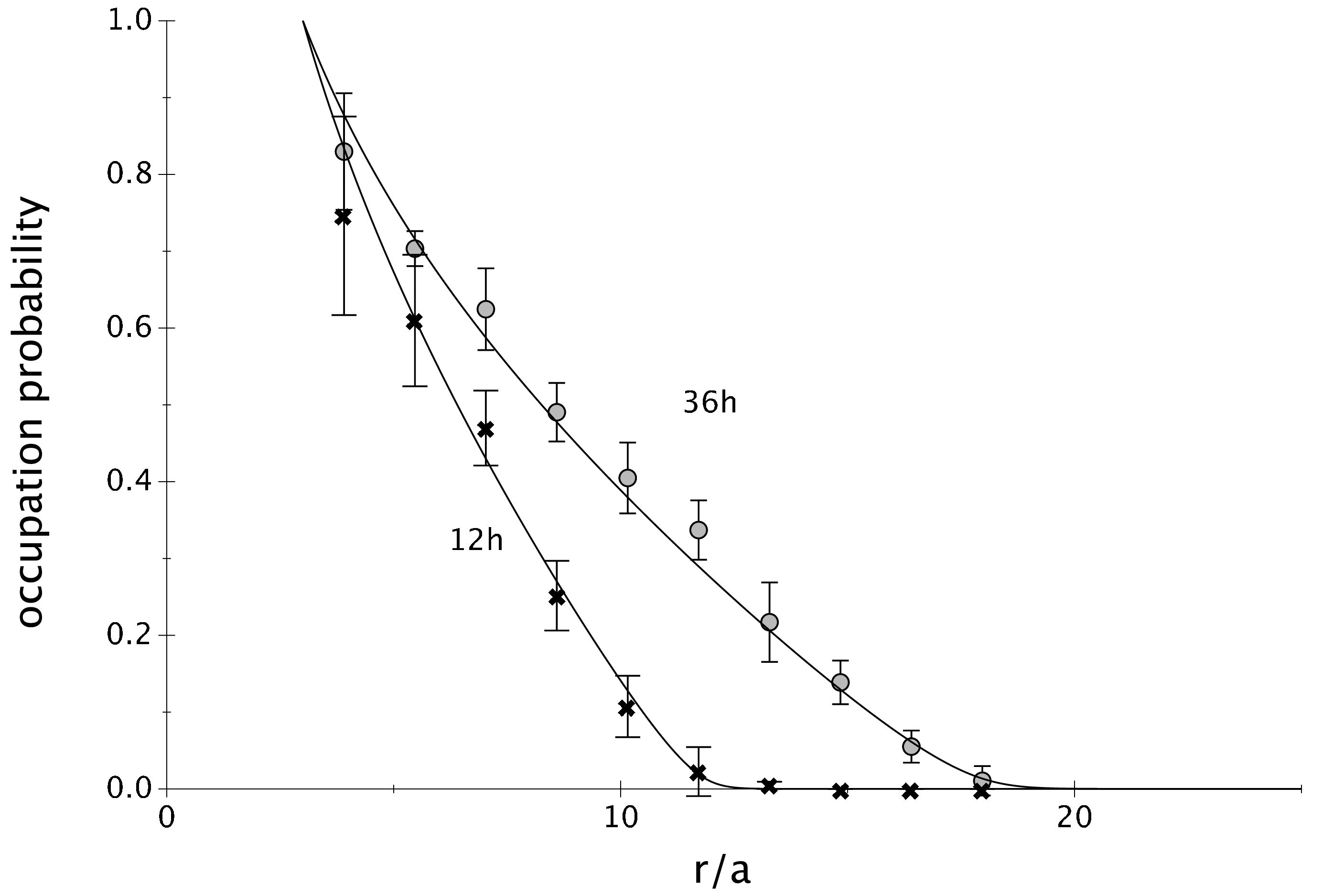}
	\end{center}
	\caption{Comparison between the numerical solution of the 
nonlinear diffusion equation (full lines) and experimental data, at 
$t=12$~h (crosses) and $t=36$~h (circles). $\Dud=1240 \pm 100~{\mu 
\mathrm{m}}^2\,\mathrm{h}^{-1}$ with $a=35.2~\mu$m.}
	\label{fig:compth_exp}
\end{figure}

This good agreement allows us to make a direct comparison between the 
numerical solution of the nonlinear diffusion equation and experimental 
cell densities. The experiments consisted in placing a spheroid of 
glioma cells (GL15) on a collagen substrate and following the evolution 
of the migration pattern during a few days. A detailed protocol has been 
described in ~\cite{aubert-et-al-migration-sans-astrocytes}. Experiments 
have been performed by Christov at the Henri Mondor Hospital. 
Figure~\ref{fig:spheroid} is a photograph of an experimental spheroid, 
36~h after cells started migrating. In order to compare the experimental 
density profiles at different times of evolution (\cf the full lines in 
Fig.~\ref{fig:compth_exp}) and the numerical solution of the diffusion 
equation, we rewrite Eq.~(\ref{eq:expression-coeff-diffusion-2d}) as

\begin{equation} 
D(\rho)=\Dud [2(1-p)+4(2p-1)\rho-2(2p-1)\rho^2],\label{eq:coeffdiff}
\end{equation}

where $\Dud$ is the value of $D$ when $p=\frac{1}{2}$ (\ie without 
interactions between cells). Note that $\Dud=\frac{1}{8}$ when the time 
step of the numerical solution coincides with that of the automaton (\cf 
Sec. III~B~2). For comparison with experimental data, the numerical time 
step is chosen equal to 1 h (and thus $\Dud$ can be different from 
$\frac{1}{8}$). The spatial length step is set to $35.2~\mu$m as in the 
automaton, which corresponds to a characteristic size of cells. The 
diffusion equation was integrated for $p=0.95$. The choice $p=1$ induces 
sharp edges at low densities that do not correspond to our experimental 
profiles.

We find that, for the coefficient $\Dud=1240 \pm 100~{\mu \mathrm{m}}^2 
\, \mathrm{h}^{-1}$, numerical solutions at 12 and 36~h agree well with 
the experimental data at the same times, \cf Fig.~\ref{fig:compth_exp}. 
One can notice that this value falls within the range of values of the 
diffusion coefficient values obtained in vivo~\cite{swanson-coeff-diff} 
(between 4 and $88~\mathrm{mm}^2/\mathrm{d}=450-10000~{\mu \mathrm{m}}^2 
\, \mathrm{h}^{-1}$), although this comparison must be taken with a 
grain of salt due to the different nature of the systems.
 
For this type of \invitro experiments, where cells are followed 
individually, the automaton remains the best way to model the system: 
time is counted as the number of cells that have left the spheroid, the 
lattice step corresponds to the mean size of a cell, and interaction 
between cells is represented by rules defined between neighboring 
cells. The only parameter to be fixed is $p$. On the contrary, for real 
tumors where the number of cells is of the order of~$10^7$, the 
automaton becomes unmanageable, and in this case, the approach through a 
diffusion equation is more efficient.

\section{Conclusion and outlook}
In this paper we set out to establish a link between a microscopic 
description of tumor cell migration with contact interactions (which we 
developed in~\cite{aubert-et-al-migration-sans-astrocytes}) and a 
macroscopic ones, based on diffusion-type equations used in most 
phenomenological models of tumor growth.

A microscopic description has obvious advantages. First, it allows a 
straightforward visualisation of the results: one ``sees'' the cells 
migrating. Moreover it can be directly compared to experiments and thus 
reveal special features of the migration like, for instance, cell 
attraction which we postulated 
in~\cite{aubert-et-al-migration-sans-astrocytes} in order to interpret 
the experimental cell distribution. It allows predictions for the 
trajectories of individual cells that can be observed in time-lapse 
microscopy 
experiments~\cite{druckenbrod-epstein-migration-enteric-neural-crest, 
young-et-al-migration-neural-crest, 
simpson-et-al-cell-scale-population-scale}. However the microscopic 
approach is not without its drawbacks. In all implementations we 
presented we have limited ourselves to a two-dimensional geometry. While 
a three-dimensional extension is, in principle, feasible it would lead 
to a substantial complication of the model. Moreover, even in a 2-D 
setting, it is not computationally feasible to accomodate the millions 
of cells present in a tumor. Our typical migration simulations involve 
at best a few thousand cells.

A macroscopic treatment on the other hand deals with mean quantities, 
like densities. Thus limitations such as the number of elementary 
entities (cells, in the present work) simply disappear. However the 
trace of microscopic interactions equally disappears unless it is 
explicitly coded into the macroscopic equations. This is a very delicate 
matter: the present work is dedicated to bridging just this gap between 
the microscopic and macroscopic approaches.

Our starting point was the microscopic description of cell migration 
which was validated through a direct comparison to experiment. Starting 
from a cellular automaton (with a given geometry and update rules) we 
have derived a \emph{nonlinear} diffusion equation by taking the 
adequate continuous limits. Our main finding was that the interactions 
between cells modify the diffusion process inducing nonlinearities.

Such nonlinearities have been observed in models of ecosystems where 
individuals tend to migrate faster in overcrowded 
regions~\cite{okubo-livre-diffusion-en-ecologie, murray-livre-tome-1}. 
This mechanism (fleeing overcrowded regions) has also been the subject 
of speculation in the context of cell 
migration~\cite{witelski-diffusion-non-lineaire}, but, in our 
experimental situation, it is irrelevant: on one hand, cells in 
overcrowded regions cannot migrate fast, even if they want to, because 
they are fully packed (thus blocked); on the other hand it has been 
shown experimentally that the nonlinearity is due to attractive 
interactions between cells (when interactions are suppressed, the 
nonlinearity diminishes)~\cite{aubert-et-al-migration-avec-astrocytes}.

While the present model is most interesting, it is fair to point out the 
precise assumptions that entered its derivation. We were based on a 
specific cellular automaton: many more do exist, the limit being only 
one's imagination. Of course for the problem at hand it was essential to 
have a geometry allowing every site to possess a sufficiently large 
number of neighbours so as not to artificially hinder the cell motion. 
Moreover the update rules were chosen {\sl in fine} through a comparison 
to experiment. But it would be interesting to see how robust the 
nonlinear diffusion coefficient is with respect to changes of the 
geometry: introducing a random deformation of the lattice or mimicking 
the migration of cancer cells on a susbtrate of astocytes in the spirit 
of~\cite{aubert-et-al-migration-avec-astrocytes}. Moreover, if one 
forgets about the application to tumor cell migration, there may exist 
several interesting cases of cellular automaton geometry and kinetic 
rules leading to nonlinear diffusion equations worth studying, in 
addition to the kinetically constrained models we have seen and that 
have their own interest for the physics of glasses. We intend to come 
back to these questions in some future publications.

Finally, an incontrovertible generalization of the present work would be 
that of three-dimensional models. While the extension of microscopic 
models to three dimensions may be computationally overwhelming, the 
solution of three-dimensional diffusion equations, be they nonlinear, 
does not present particular difficulties. Thus once the automaton rules 
(mimicking the cell-cell interaction) have been used for the derivation 
of a diffusion equation like Eq.~(\ref{eq:edp-rho}) with 
coefficient~(\ref{eq:expression-coeff-diffusion-3d}), we can proceed to 
the study of the macroscopic model, coupling diffusion to proliferation 
(plus, perhaps, other macroscopically described effects) for the 
modeling of real-life tumors (in particular, glioblastomas). This is a 
path we intend to explore in some future work.

\begin{acknowledgments}
 We acknowledge financial support from the ``Comit\'e de financement 
des th\'eoriciens'' of the IN2P3 (CNRS).
\end{acknowledgments}

\def\pl{Phys.\ Lett.\xspace}
\def\pra{Phys.\ Rev.\ A\xspace}
\def\prb{Phys.\ Rev.\ B\xspace}
\def\prc{Phys.\ Rev.\ C\xspace}
\def\prd{Phys.\ Rev.\ D\xspace}
\def\pre{Phys.\ Rev.\ E\xspace}
\def\prl{Phys.\ Rev.\ Lett.\xspace}
\def\rmp{Rev.\ Mod.\ Phys.\xspace}

\newcommand{\jphysa}{J. Phys. A -- Math. Gen.}
\newcommand{\jphysc}{J. Phys. C -- Solid State Phys.}
\newcommand{\jstatphys}{J. Stat. Phys.}
\newcommand{\physrev}{Phys. Rev.}
\newcommand{\physrep}{Phys. Rep.}
\newcommand{\zpb}{Z. Phys. B}
\newcommand{\zphys}{Z. Phys.}
\newcommand{\theorcompsci}{Theor. Comp. Sci.}
\newcommand{\epl}{Europhys. Lett.}
\newcommand{\epjb}{Eur. Phys. J. B}
\newcommand{\epje}{Eur. Phys. J. E}
\newcommand{\amai}{Annals of Mathematics and Artificial Intelligence}
\newcommand{\jmathphys}{J. Math. Phys.}
\newcommand{\communmathphys}{Commun. Math. Phys.}
\newcommand{\physlett}{Phys. Lett.}
\newcommand{\revmodphys}{\rmp}
\newcommand{\theormathphys}{Theor. Math. Phys.}
\newcommand{\jstatmech}{J. Stat. Mech.}
\newcommand{\advphys}{Adv. Phys.}
\newcommand{\rsa}{Random Structures Algorithms}
\newcommand{\annprob}{Annals of Probability}
\newcommand{\tcs}{Theor. Comp. Sci}
\newcommand{\jhep}{J. High Energy Phys.}
\newcommand{\jphyscondmat}{J. Phys. Cond. Mat.}
\newcommand{\ssc}{Solid State Commun.}
\newcommand{\zetf}{Zh. Eksp. Teor. Fiz.}
\newcommand{\jetp}{Sov. Phys. JETP}
\newcommand{\jap}{J. Appl. Prob.}
\newcommand{\jrsocinterface}{J. R. Soc. Interface}
\newcommand{\physbiol}{Phys. Biol.}
\newcommand{\jpc}{J. Phys. Chem.}
\newcommand{\philtransrsoca}{Phil. Trans. R. Soc. A}
\newcommand{\jmathbiol}{J. Math. Biol.}
\newcommand{\bullmathbiol}{Bull. Math. Biol.}
\newcommand{\mathmodmethapplsci}{Math. Mod. Meth. Appl. Sci.}
\newcommand{\siamjapplmath}{SIAM J. Appl. Math.}

\newcommand{\annphysny}{Ann. Phys. (N.Y.)}
\newcommand{\applphys}{Appl. Phys.}

\bibliography{micromacro}

\end{document}